\documentclass[]{paper}

\usepackage[utf8]{inputenc}

\usepackage{fullpage}

\usepackage{times}
\usepackage[small,compact]{titlesec}

\usepackage{boxedminipage}
\usepackage{url}

\usepackage{listings}

\usepackage{epsfig}
\usepackage{ifpdf}
\usepackage{amsmath}
\usepackage{color}

\usepackage{listings}
\definecolor{commentcolor}{rgb}{0,0.5,0.2}
\usepackage{graphicx}

\newcommand{\ILCE}[1]{\lstinline[language=c++,breaklines=true,basicstyle=\ttfamily]@#1 @}

\title{OCR extensions -- local identifiers, labeled GUIDs, file IO, and data block partitioning}

\author{Jiri Dokulil, Siegfried Benkner\\Research Group Scientific Computing\\University of Vienna\\(jiri.dokulil,siegfried.benkner)@univie.ac.at}

\begin{document}

\maketitle

\begin{abstract}
We present several proposals for extending the Open Community Runtime (OCR) specification. The extension are identifiers with local validity, which use the concept of futures to provide OCR implementations more optimization opportunities, labeled GUIDs with creator functions, which are based on the local identifiers and allow the developer to create arrays of OCR objects that are safe from race conditions in case of concurrent creation of objects, a simple file IO interface, which builds on top of the existing data block concepts, and finally data block partitioning, which allows better control and flexibility in situations where multiple tasks want to access disjoint parts of a data block.
\end{abstract}

\section{Introduction}

The Open Community Runtime specification \cite{MattsonOCR15} is still in an early stage of development. There are many opportunities to extend the API, in order to provide it with new functionality or to provide more efficient ways of building OCR applications. In the following text, we present several proposals for such extensions. The extensions have been implemented in the experimental OCR-V1 single-threaded OCR runtime, available here:\\\url{http://homepage.univie.ac.at/jiri.dokulil/projects/}

\paragraph{Status of the text} This text describes the current state of a work-in-progress and will be updated in the future.
\\

\section{Constraints}

In our considerations, we do not target a specific OCR implementation, but we do make some assumptions about what the implementation looks like.

\paragraph{Messages}
We assume that the runtime works by sending messages between the nodes of a distributed system. A call to the OCR API typically translates to one or more messages, that could either be handled locally or sent to another node for processing.

\paragraph{GUIDs}
The OCR specification does not prescribe how GUIDs should look like. However, we assume that a GUID may encode information that is not available before the call to the OCR API which creates the object identified by the GUID. For example, the GUID may encode the type of the object, or some of the arguments of the creation call, like the number of pre-slots of a task. It is therefore not possible to pre-allocate GUIDs. Furthermore, we assume that it may not be possible to create GUIDs locally. It may be necessary to communicate with another node in order to discover the actual value of the GUID. Possible examples of GUIDs under these contraints include:

\begin{itemize}
\item Address of the OCR object in a global (shared) address space.
\item A pair consisting of a node identifier and sequential identifier of the object within the node.
\item A tuple consisting of a node identifier, sequential identifier, and object type (enumeration).
\item Global sequential identifier of the object, maintained by a coordinator node.
\end{itemize}

\section{Local identifiers}

Ideally, it should be possible to process all OCR API calls (made by tasks) locally, without the need to send a message to a remote node and wait for the reply. With relaxed requirements for error detection, this is possible for most of the API calls. For example, the \ILCE{ocrEventSatisfy} call can be handled by a message that would inform the event that its pre-slot has been satisfied. The call only affects the satisfied event, not the calling task. Therefore, the \ILCE{ocrEventSatisfy} function can return immediately, without waiting for the message to be processed.

Functions which create objects are an exception. They have to return the GUID of the newly created object. With the restrictions we have placed on the GUIDs, some implementations may need to contact another node, for example to get the current value of the ID sequence of the node where the object will actually be created. Once the GUID is determined and returned to the calling task, it may for example store the GUID in a data block, so that other tasks can read it and use it in further OCR API calls. However, it is also common that the GUID is only used to make several calls to the OCR API by the task that created the object. The GUID is not stored in any data block. A typical example would be when a task is created and all of its dependences are immediately defined by the creating task.

If the GUID is not saved, it is not necessary to have a true GUID -- a \emph{globally} unique identifier. Therefore, an identifier with a local validity may be sufficient. Since the OCR abstracts nodes, the only applicable scope besides the global scope is a task. Therefore, the OCR call which creates an object may return a local identifier (a LID), which may only be used to make OCR API calls by the task that created the object. Such a LID cannot be saved to a data block and used by another task.

A different view of the LIDs is to view them as futures. Instead of returning the new object's GUID, a future is returned. The future may be used to make further OCR API calls, but the runtime does not yet send out messages that implement these calls. At some later point in time (possibly after the task has finished) the future will be resolved, allowing the runtime to  update the messages, providing the actual GUID instead of the future.

Consider the following example:
\begin{lstlisting}
void launch_task(ocrGuid_t task_template, ocrGuid_t data)
{
  ocrGuid_t task_lid;
  ocrEdtCreate(&task_lid,task_template,0,0,1,0,EDT_PROP_LID,NULL_GUID,0);
  ocrAddDependency(data,task_lid,0,DB_MODE_EW);
}
\end{lstlisting}

This may be implemented like this:
\begin{enumerate}
\item \ILCE{ocrEdtCreate} generates a new LID $L_1$, sends out a message $M_{create}$ specifying LID $L_1$ for the object. It immediately returns, providing $L_1$ as the task's identifier.
\item \ILCE{ocrAddDependency} generates a new message $M_{dep}$ with the data block's GUID $G_{data}$ and task's LID $L_1$. It returns immediately.
\item At any point in the future, the $M_{create}$ message is processed. The EDT is created with GUID $G_1$. A new message $M_{map}$ is generated, describing the mapping from $L_1$ to $G_1$. This message is sent back to the node that made the original \ILCE{ocrEdtCreate} call.
\item $M_{map}$ is received and processed. As a result the $M_{dep}$ is updated, replacing $L_1$ with $G_1$. Then, it is submitted for processing as it would be normally.
\end{enumerate}

To make the API more usable, the LID extension should not just provide extra flags for the calls of function that create OCR objects. It should also include at least the two following functions:
\begin{lstlisting}
enum ocrIdType
{
  OCR_ID_GUID, // ID is GUID
  OCR_ID_LID, // ID is LID
  OCR_ID_UNK, // Error, incorrect ID
};
ocrIdType ocrGetIdType(ocrGuid_t id);
u8 ocrGetGuid(ocrGuid_t* guid, ocrGuid_t sourceId);
\end{lstlisting}

The \ILCE{ocrGetGuid} would be an exception to the ``no blocking'' rule in the OCR API. It blocks until the GUID can be obtained (until the $M_{map}$ message from the previous example arrives). The \ILCE{ocrGetIdType} may be important for optimization of the application code. In some situations, the runtime may be able to return a real GUID, rather than LID, even without communication. Therefore the GUID may be returned, even if a LID was requested. This is not a problem, since the GUID can be used everywhere instead of a LID. But the application may be able to use the fact that it received a GUID, even if the LID would be sufficient for now.

\section{Labeled GUIDs}

The OCR specification draft 1.0.1 contains a proposal for labeled GUIDs. The idea is to provide a convenient way of managing a large number of GUIDs without the need to pass all such GUIDs from task to task. Instead, a map is created. The user can then obtain a GUID from the map just by providing a GUID of the map and coordinates of the object within the map. There are two issues connected to creation of objects in such maps. First, we assume that it may not be possible to create a GUID before the parameters of the creation call are known. Second, the map should be able to deal with concurrent requests to acquire the same object from the map. Since it should be possible to create and destroy objects in the map as needed, not just when the map itself is created and destroyed, the map should take care of all necessary synchronization and make sure that objects are created only once.

We propose an alternative API, based on the LIDs introduced in the previous section. The purpose of the proposal is to provide the highest level of guarantee in the case of concurrent object creation (\ILCE{GUID_PROP_BLOCK} in the 1.0.1 OCR proposal), without the need to block the task's execution.

\begin{lstlisting}
typedef void(*ocrCreator_t)(ocrGuid_t objectLid, u64 index, u32 paramc, u64* paramv, u32 guidc, ocrGuid_t* guidv);

u8 ocrMapCreate(ocrGuid_t* mapGuid, u64 size, ocrCreator_t creatorFunc, u32 paramc, u64* paramv, u32 guidc, ocrGuid_t* guidv);
u8 ocrMapDestroy(ocrGuid_t mapGuid);
u8 ocrMapGet(ocrGuid_t* lid, ocrGuid_t mapGuid, u64 index);
\end{lstlisting}

The \ILCE{ocrMapCreate} function creates a new map, containing \ILCE{size} objects, which are not yet initialized. The user specifies a \ILCE{creatorFunc} -- a \emph{creator} function responsible for creation of objects in the map. The \ILCE{paramv} and \ILCE{guidv} arrays are forwarded to the creator function. Initially, the map is empty and no object is created. A task may then call \ILCE{ocrMapGet} to get an object at a specified index from the map. If the object does not exist, the map calls the creator function and the object is created. However, this creation has to be properly synchronized by the runtime, to make sure that concurrent calls to \ILCE{ocrMapGet} with the same index return the same object. At the same time, we would like all OCR API calls to return immediately, without any communication. Since synchronization without communication is not possible, we use LIDs to solve the problem. The \ILCE{ocrMapGet} returns a LID of the object. This can be done locally. If two tasks make an identical \ILCE{ocrMapGet} call, they get different LIDs, but once resolved, the LIDs will point to the same object (same GUID).

The remaining issue is the inside of the creator function. We have to consider two different options. First, the runtime may prefer to use a sequence of consecutive GUIDs for the map. In this case, it needs to set up all GUIDs in the map at the time the map is created. Second, the runtime may only be able to create the GUID of an object when the object is created. Therefore, the GUIDs in the map are only filled as the individual objects get created. We propose to also use LIDs in this case. However, they serve a different purpose. When a map needs to create an object, it first creates a LID. The LID is then passed to the creator function, which in turn passes it to the call of the \ILCE{ocrXxxCreate} function. This changes the GUID parameter of such function from \ILCE{out} parameter to \ILCE{inout}. We also propose to add a new flag to such functions, to distinguish the situation.

Following code is an example application using our labeled GUID API to create a 2D matrix of tasks, where each task depends on its immediate neighbors in the up and left directions.

\begin{lstlisting}
void creator(ocrGuid_t objectLid, u64 index, u32, u64* paramv, u32, ocrGuid_t* guidv) {
  u64 width = paramv[0];
  u64 height = paramv[1];
  u64 x = index % width;
  u64 y = index / width;
  u64 params[] = { x, y };
  ocrGuid_t deps[] = { guidv[0], UNINITIALIZED_GUID, UNINITIALIZED_GUID };
  if (x == 0) deps[1] = NULL_GUID;
  if (y == 0) deps[2] = NULL_GUID;
  ocrEdtCreate(&objectLid, guidv[1], EDT_PARAM_DEF, params, EDT_PARAM_DEF, deps, EDT_PROP_MAPPED, NULL_GUID, 0);
  //The objectLid should now contain a valid GUID, but we don't need it here.
}

struct data {
  ocrGuid_t map;
  ocrGuid_t task_template;
  u64 width;
  u64 height;
};

u64 coords(data* data, u64 x, u64 y) {
  return x + y * data->width;
}

ocrGuid_t work(u32 paramc, u64* paramv, u32 depc, ocrEdtDep_t depv[]) {
  u64 x = paramv[0];
  u64 y = paramv[1];
  data* data = (data*)depv[0].ptr;
  if (x == data->width - 1 && y == data->height - 1)
  {
    //The last item, we are done
    ocrEdtTemplateDestroy(data->task_template);
    ocrMapDestroy(data->map);
    ocrDbDestroy(depv[0].guid);
    ocrShutdown();
    return NULL_GUID;
  }
  //Do the work here.
  //...
  //Done.
  if (x < data->width - 1) {
    //Not the last column, satisfy preslot of the task to the right
    ocrGuid_t task;
    ocrMapGet(&task, data->map, coords(data, x + 1, y));
    ocrAddDependence(NULL_GUID, task, 1, DB_DEFAULT_MODE);
  }
  if (y < data->height - 1) {
    //Not the last row, satisfy preslot of the task below
    ocrGuid_t task;
    ocrMapGet(&task, data->map, coords(data, x, y + 1));
    ocrAddDependence(NULL_GUID, task, 2, DB_DEFAULT_MODE);
  }
  return NULL_GUID;
}

ocrGuid_t mainEdt(u32 paramc, u64* paramv, u32 depc, ocrEdtDep_t depv[]) {
  ocrGuid_t db, task;
  void* ptr;
  ocrDbCreate(&db, &ptr, sizeof(data), 0, NULL_GUID, NO_ALLOC);
  data* data = (data*)ptr;
  data->width = 3;
  data->height = 3;
  ocrEdtTemplateCreate(&data->task_template, work, 2, 3);
  u64 params[] = { data->width, data->height };
  ocrGuid_t guids[] = { db, data->task_template };
  ocrMapCreate(&data->map, data->height * data->width, &creator, 2, params, 2, guids);
  ocrMapGet(&task, data->map, 0);
  return NULL_GUID;
}
\end{lstlisting}

\paragraph{Note on data blocks} Unlike making direct calls to the \ILCE{ocrDbCreate} function from the task, the creator-based API does not allow the pointer to a data block to be returned to the caller. However, if we want to deal with concurrent calls to \ILCE{ocrDbCreate}, the pointer may not be returned anyway, since the call may be made on two different nodes, making it impossible to provide a valid pointer on both. With the creator, only a LID is returned. To access the data, a new task has to be created and have the LID as the data on one of its pre-slots. This gives the runtime the opportunity to deal with the situation.

\section{File IO}

Dealing with files is still an open question in the OCR. The traditional \ILCE{fopen}-based approach does not mix well with the resilience requirements of the OCR, since direct access to files allows tasks to have side effects outside of the OCR. To provide at least the most basic file IO, we propose to add ``file-mapped data blocks''. These are data blocks whose contents are read from a file and then saved back to the file. The proposed API looks like this:

\begin{lstlisting}
u8 ocrFileOpen(ocrGuid_t* fileGuid, const char* path, const char* mode, ocrGuid_t* descriptorDb, u32 properties);
u8 ocrFileRelease(ocrGuid_t fileGuid);
ocrGuid_t ocrFileGetGuid(void* descriptor);
u64 ocrFileGetSize(void* descriptor);
u8 ocrFileGetChunk(ocrGuid_t* chunkDbGuid, ocrGuid_t fileGuid, u64 offset, u64 size);
\end{lstlisting}

The \ILCE{ocrFileOpen} function returns two GUIDs (the second is optional). The GUID of the file and a GUID which represents a data block, which contains information about the file. This data block cannot be used directly (no pointer is provided). This is intentional, since it may take some time to open the file and find out its properties. To use the information, a task needs to receive the data block via one of its pre-slots. This allows the runtime to delay the task until the information about the file is available. Once the task starts, it may use \ILCE{ocrFileGetSize} to find out the size of the file when it was opened. The  \ILCE{ocrFileGetGuid} function is a convenience for a typical scenario, where one task calls \ILCE{ocrFileOpen} and passes the \ILCE{descriptorDb} to another task. The second task then reads the size of the file and creates other tasks to process it. To do that, the second tasks needs to know the GUID of the file. Without \ILCE{ocrFileGetGuid}, a data block would have to be created to send the GUID from the first to the second task.

Once a file has been opened, the \ILCE{ocrFileGetChunk} call can be used to map a contiguous part of the file into a data block. If the data block (\ILCE{offset}+\ILCE{size}) extends beyond the end of the file and the file is writeable, it is enlarged. It is not allowed to define overlapping chunks. The runtime may decide not to write the chunk data back to the file if the chunk was not used in EW or RW mode by a task. However, if the block would cause the file to be enlarged, it has to be enlarged, even if the data is not written.

The following example reads a file, which contains 32bit unsigned integers, multiplies each value by 2 and saves the updated values. Two tasks are used to perform multiplication in parallel.

\begin{lstlisting}
ocrGuid_t work(u32 paramc, u64* paramv, u32 depc, ocrEdtDep_t depv[]) {
  u32* data = (u32*)depv[0].ptr;
  for (std::size_t i = 0; i < paramv[0]; ++i) {
    data[i] *= 2;
  }
  ocrDbDestroy(depv[0].guid);
  return NULL_GUID;
}

ocrGuid_t check(u32 paramc, u64* paramv, u32 depc, ocrEdtDep_t depv[]) {
  ocrGuid_t chunk1, chunk2, worker1, worker2, worker_template, finish_template, finish_task, event1, event2;
  u64 size = ocrFileGetSize(depv[0].ptr);
  ocrFileGetChunk(&chunk1, ocrFileGetGuid(depv[0].ptr), 0, size/2);
  ocrFileGetChunk(&chunk2, ocrFileGetGuid(depv[0].ptr), size/2, size/2);
  ocrFileRelease(ocrFileGetGuid(depv[0].ptr));
  ocrDbDestroy(depv[0].guid);
  ocrEdtTemplateCreate(&worker_template, work, 1, 1);
  ocrEdtTemplateCreate(&finish_template, test8_finish, 0, 2);
  u64 params[] = { (size / sizeof(u32))/2 };
  ocrEdtCreate(&worker1, worker_template, EDT_PARAM_DEF, params, EDT_PARAM_DEF, 0, 0, NULL_GUID, &event1);
  ocrEdtCreate(&worker2, worker_template, EDT_PARAM_DEF, params, EDT_PARAM_DEF, 0, 0, NULL_GUID, &event2);
  ocrEdtCreate(&finish_task, finish_template, EDT_PARAM_DEF, 0, EDT_PARAM_DEF, 0, 0, NULL_GUID, 0);
  ocrAddDependence(event1, finish_task, 0, DB_DEFAULT_MODE);
  ocrAddDependence(event2, finish_task, 1, DB_DEFAULT_MODE);
  ocrAddDependence(chunk1, worker1, 0, DB_MODE_RO);
  ocrAddDependence(chunk2, worker2, 0, DB_MODE_RO);
  ocrEdtTemplateDestroy(worker_template);
  ocrEdtTemplateDestroy(finish_template);
  return NULL_GUID;
}

ocrGuid_t mainEdt(u32 paramc, u64* paramv, u32 depc, ocrEdtDep_t depv[]) {
  ocrGuid_t info, file, task, checker_template;
  ocrFileOpen(&file, "data.dat", "rb+", &info, 0);
  ocrEdtTemplateCreate(&checker_template, check, 0, 1);
  ocrGuid_t deps[] = { info };
  ocrEdtCreate(&task, checker_template, EDT_PARAM_DEF, 0, EDT_PARAM_DEF, deps, 0, NULL_GUID, 0);
  ocrEdtTemplateDestroy(checker_template);
  return NULL_GUID;
}
\end{lstlisting}

\section{Data block partitioning}

Currently, if two tasks want to modify two parts of the same data block, they both need to acquire it in RW mode. The OCR runtime has to assume that the tasks may also share data through the data block by reading and writing to the same position in the data block, a situation similar to \emph{false sharing} of cache lines. It may be more efficient to explicitly represent the fact, that these two tasks don't access overlapping parts of the data block. At the moment, there is no way of expressing this directly using the OCR API. The data block could be split into two (or more) data blocks, providing each of the two tasks with one smaller block. The tasks can acquire their respective data blocks in EW mode, without preventing parallel execution. But having the data split into two parts may be inconvenient for other parts of the code, where two GUIDs would have to be passed instead of one. Or it may be necessary to dynamically change the way the data is split into blocks.

\subsection{Higher-level model}

The ultimate goal is to have a higher-level programming model (HLP) on top of the OCR. The HLP will be responsible for management of the data blocks, so it may be able to create the data blocks in such a way that there is no false sharing of the data blocks. It may be necessary to use different partition strategies on one data block. For example, given a data block with four items $A$, $B$, $C$, and $D$ we may first need partitions $p_1=\{A,B\}$ and $p_2=\{C,D\}$, but later partitions $q_1=\{A\}$ and $q_2=\{B,C,D\}$. There are two basic options. First, creating three data blocks ($b_1=\{A\}$, $b_2=\{B\}$, and $b_3=\{C,D\}$) and providing multiple data blocks for each partition: $p_1=b_1\cup b_2$, $p_2=b_3$; $q_1=b_1$, $q_2=b_2\cup b_3$. This complicates implementation of tasks that process the partitions, since they need to accept data that is split into multiple data blocks. The second option is to create new data blocks for partitions $p_1$ and $p_2$ and then copy the data into two other blocks for partitions $q_1$ and $q_2$. This requires extra memory and data movement. Due to the issues present in both options, it may be interesting to investigate alternatives where the partitioning is made explicit using the OCR API.

\subsection{Explicit DB partitioning support in the OCR API}

An alternative would be to add explicit partitioning support to the OCR:

\begin{lstlisting}
typedef struct {
  u64 offset;//[in] the offset of the partition within the parent block
  u64 size;//[in] size of the partition
  ocrGuid_t guid;//[out] GUID of the new data block
} ocrDbPart_t;
#define OCR_DB_PARTITION_STATIC ((u16) 0x1)
u8 ocrDbPartition(ocrGuid_t dbGuid, u32 partCount, ocrDbPart_t* partitions, u32 properties);
\end{lstlisting}

The \ILCE{ocrDbPartition} function accepts a list of partition definitions and returns a list of GUIDs for the corresponding partitions. The partitions are not allowed to overlap, but there can be ``holes'' -- sections of the data block that don't belong to any partition. Unless \ILCE{OCR_DB_PARTITION_STATIC} is provided, \ILCE{ocrDbPartition} can be used again on the same data block to create more partitions. However, no partitions (both new and old) may overlap. If partitioning is static, it's not possible to use \ILCE{ocrDbPartition} on the data block until all partitions have been destroyed using \ILCE{ocrDbDestroy}.

It is disallowed to use a partitioned data block to satisfy a pre-slot (either directly on indirectly via a dependence) of a task that may become runnable before all partitions have been destroyed. A more relaxed condition would be to allow the data block to be used this way, but the runtime may decide not to allow any task to actually acquire the data block until all partitions have been released. It is still not possible to pass the partitioned data block and one of its partitions to the same task, since that may cause a deadlock.

It is possible to further partition data blocks acquired by partitioning. Recursive partitioning is allowed, with the same constraints. It is not possible to create a partition that does not represent a single contiguous block of the partitioned DB. So, for example, it's not possible to create cyclic partitions. There is also no API to query the partition information of a data block or to check whether a data block is a regular data block, partitioned data block, or a partition.

The following example demonstrates the partitioning API. It creates a data block, splits it into two partitions, executes a worker task on each partition, and finally uses the original data block in the final task.

\begin{lstlisting}
ocrGuid_t finish(u32 paramc, u64* paramv, u32 depc, ocrEdtDep_t depv[]) {
  u64 sum = 0;
  u32* data = (u32*)depv[0].ptr;
  for (std::size_t i = 0; i < 1024; ++i)  sum+=data[i];
  PRINTF("%lu\n", sum);
  ocrDbDestroy(depv[0].guid);
  ocrShutdown();
  return NULL_GUID;
}

ocrGuid_t work(u32 paramc, u64* paramv, u32 depc, ocrEdtDep_t depv[]) {
  u32* data = (u32*)depv[0].ptr;
  for (std::size_t i = 0; i < 512; ++i) {
    data[i] *= paramv[0];
  }
  ocrDbDestroy(depv[0].guid);
  return NULL_GUID;
}

ocrGuid_t mainEdt(u32 paramc, u64* paramv, u32 depc, ocrEdtDep_t depv[]) {
  ocrGuid_t block, worker_template, worker1, worker1_event, worker2, worker2_event, finish_template, finish_task;
  void* ptr;
  ocrDbCreate(&block, &ptr, 1024*sizeof(u32), 0, NULL_GUID, NO_ALLOC);
  u32* data = (u32*)ptr;
  for (u32 i = 0; i < 1024; ++i) {
    data[i] = 1;// i + 1;
  }
  ocrDbPart_t parts[2];
  parts[0].offset = 0;
  parts[0].size = 512 * sizeof(u32);
  parts[1].offset = 512 * sizeof(u32);
  parts[1].size = 512 * sizeof(u32);
  ocrDbRelease(block);//could also come later, but the data is no longer needed
  ocrDbPartition(block, 2, parts, 0);
  ocrEdtTemplateCreate(&worker_template,   work, 1, 1);
  ocrEdtTemplateCreate(&finish_template,   finish, 0, 3);
  u64 params1[] = { 2 };
  u64 params2[] = { 6 };
  ocrEdtCreate(&finish_task, finish_template, EDT_PARAM_DEF, 0, EDT_PARAM_DEF, 0,0, NULL_GUID,0);
  ocrEdtCreate(&worker1, worker_template, EDT_PARAM_DEF, params1, EDT_PARAM_DEF, 0, 0, NULL_GUID, &worker1_event);
  ocrEdtCreate(&worker2, worker_template, EDT_PARAM_DEF, params2, EDT_PARAM_DEF, 0, 0, NULL_GUID, &worker2_event);
  ocrEdtTemplateDestroy(worker_template);
  ocrEdtTemplateDestroy(finish_template);
  ocrAddDependence(block, finish_task, 0, DB_MODE_RO);
  ocrAddDependence(worker1_event, finish_task, 1, DB_DEFAULT_MODE);
  ocrAddDependence(worker2_event, finish_task, 2, DB_DEFAULT_MODE);
  ocrAddDependence(parts[0].guid, worker1, 0, DB_MODE_EW);
  ocrAddDependence(parts[1].guid, worker2, 0, DB_MODE_EW);
  return NULL_GUID;
}
\end{lstlisting}

\subsection{Paritioning as an optimization of data block copying (using \ILCE{ocrDbCopy})}

The 0.9 version of the OCR proposal contained a function to copy data blocks in an asynchronous way:

\begin{lstlisting}
u8 ocrDbCopy(ocrGuid_t destination, u64 destinationOffset, ocrGuid_t source, u64 sourceOffset, u64 size, u64 copyType, ocrGuid_t * completionEvt);
\end{lstlisting}

This could be used, for example by the higher-level programming model, to copy the original data block into several data blocks that represent the partitions. In certain circumstances (if the \ILCE{destinationOffset} is 0 and \ILCE{size} covers the whole destination data block), the runtime could view it as creation of a partition, as if the special API from the previous section was used. It could even skip actually copying the memory, reusing the source buffer, with copy-on-write semantics. That could still mean that some of the data will be copied, but it still gives the runtime some room for optimizations. Furthermore, the \ILCE{copyType} argument could be used to indicate that the source data block will not be used by any task and the data will be copied back before the destination block is destroyed, providing functionality similar to \ILCE{ocrDbPartition}, with partitioning that is potentially zero-copy. Naturally, it is also not possible to use the partitioned data block before partitions have been copied back. The user is responsible for ensuring that this rule is observed.

Following example demonstrates how to achieve partitioning similar to the example of the explicit API:

\begin{lstlisting}
ocrGuid_t finish(u32 paramc, u64* paramv, u32 depc, ocrEdtDep_t depv[]) {
  ocrDbDestroy(depv[3].guid);//destroy the "params" data block
  u64 sum = 0;
  u32* data = (u32*)depv[0].ptr;
  for (std::size_t i = 0; i < 1024; ++i) {
  {
    sum += data[i];
  }
  PRINTF("%lu\n", sum);
  ocrDbDestroy(depv[0].guid);
  ocrShutdown();
  return NULL_GUID;
}

ocrGuid_t work(u32 paramc, u64* paramv, u32 depc, ocrEdtDep_t depv[]) {
  u32* data = (u32*)depv[0].ptr;
  ocrGuid_t* guids = (ocrGuid_t*)depv[1].ptr;
  for (std::size_t i = 0; i < 512; ++i) {
    data[i] *= paramv[0];
  }
  ocrDbRelease(depv[0].guid);
  ocrGuid_t event;
  ocrDbCopy(guids[1], paramv[2] * sizeof(u32), depv[0].guid, 0, 512 * sizeof(u32), DB_COPY_PARTITION_BACK, &event);//DB_COPY_PARTITION_BACK entails destruction of the source
  ocrAddDependence(event, guids[0], paramv[1], DB_MODE_NULL);
  return NULL_GUID;
}

ocrGuid_t mainEdt(u32 paramc, u64* paramv, u32 depc, ocrEdtDep_t depv[]) {
  ocrGuid_t block, params, worker_template, worker1, worker2, finish_template, finish_task, chunk1, chunk2, chunk1_copied, chunk2_copied;
  void *ptr, *params_ptr;
  ocrDbCreate(&block, &ptr, 1024 * sizeof(u32), 0, NULL_GUID, NO_ALLOC);
  u32* data = (u32*)ptr;
  for (u32 i = 0; i < 1024; ++i) {
    data[i] = 1;// i + 1;
  }
  ocrDbRelease(block);//could also come later, but the data is no longer needed
  ocrDbCreate(&params, &params_ptr, 2 * sizeof(ocrGuid_t), 0, NULL_GUID, NO_ALLOC);
  ocrDbCreate(&chunk1, &ptr, 512 * sizeof(u32), DB_PROP_NO_ACQUIRE, NULL_GUID, NO_ALLOC);//with DB_PROP_NO_ACQUIRE, the runtime may decide not to allocate any memory at this time
  ocrDbCreate(&chunk2, &ptr, 512 * sizeof(u32), DB_PROP_NO_ACQUIRE, NULL_GUID, NO_ALLOC);
  ocrDbCopy(chunk1, 0, block, 0, 512 * sizeof(u32), DB_COPY_PARTITION, &chunk1_copied);
  ocrDbCopy(chunk2, 0, block, 512 * sizeof(u32), 512 * sizeof(u32), DB_COPY_PARTITION, &chunk2_copied);
  ocrEdtTemplateCreate(&worker_template, work, 3, 2);
  ocrEdtTemplateCreate(&finish_template, finish, 0, 4);
  u64 params1[] = { 2, 1, 0 };
  u64 params2[] = { 6, 2, 512 };
  ocrEdtCreate(&finish_task, finish_template, EDT_PARAM_DEF, 0, EDT_PARAM_DEF, 0, 0, NULL_GUID, 0);
  ocrEdtCreate(&worker1, worker_template, EDT_PARAM_DEF, params1, EDT_PARAM_DEF, 0, 0, NULL_GUID, 0);
  ocrEdtCreate(&worker2, worker_template, EDT_PARAM_DEF, params2, EDT_PARAM_DEF, 0, 0, NULL_GUID, 0);
  ((ocrGuid_t*)params_ptr)[0] = finish_task;
  ((ocrGuid_t*)params_ptr)[1] = block;
  ocrDbRelease(params);
  ocrEdtTemplateDestroy(worker_template);
  ocrEdtTemplateDestroy(finish_template);
  ocrAddDependence(block, finish_task, 0, DB_MODE_RO);
  ocrAddDependence(params, finish_task, 3, DB_MODE_RO);
  ocrAddDependence(chunk1_copied, worker1, 0, DB_MODE_EW);
  ocrAddDependence(chunk2_copied, worker2, 0, DB_MODE_EW);
  ocrAddDependence(params, worker1, 1, DB_MODE_CONST);
  ocrAddDependence(params, worker2, 1, DB_MODE_CONST);
  return NULL_GUID;
}
\end{lstlisting}

Since the data block is created with \ILCE{DB_PROP_NO_ACQUIRE} and then filled by calling \ILCE{ocrDbCopy} with \ILCE{DB_COPY_PARTITION}, the runtime knows not to allocate memory for the data block and also that the block is in fact a partition of another block. As a result, it can reuse the memory of the partitioned block. On the other hand, if an OCR implementation does not want to provide partitioning, it only needs to ignore \ILCE{DB_COPY_PARTITION} and extend \ILCE{ocrDbCopy} to release the source data block when \ILCE{DB_COPY_PARTITION_BACK} is used. This makes the extension relatively cheap to implement if the zero-copy functionality is not desired.


\end{document}